\documentclass[usenatbib]{aa}

\usepackage{epsfig} 
\usepackage{graphicx}
\usepackage{amssymb}
\usepackage{natbib}

\def\inte{{\em INTEGRAL}}
\def\XMM{{\em XMM-Newton}}

\def\chan{{\em Chandra}}

\def\beppo{{\em BeppoSAX}}
\def\rxte{{\em RXTE}}
\def\swift{{\em Swift}}
\def\src{Swift\,J1749.4$-$2807}

\begin{document}

\title{\inte,\ \swift,\ and \rxte\ observations of the 518~Hz accreting transient pulsar Swift\,J1749.4$-$2807}

\author{C. Ferrigno  
	\inst{1}
	\and
	E. Bozzo 
	\inst{1}
	\and 
	M. Falanga
	\inst{2}
	\and
	L. Stella
	\inst{3}
	\and
	S. Campana
	\inst{4}
	\and 
	T. Belloni
	\inst{4}
	\and
	G. L. Israel
	\inst{3}
	\and
	L. Pavan
	\inst{1}
	\and
	E. Kuulkers
	\inst{5}
	\and
	A. Papitto
	\inst{6,7}
}

\authorrunning{C. Ferrigno et al.}
  \titlerunning{X-ray observations of Swift\,J1749.4-2807}
  \offprints{C. Ferrigno}

\institute{ISDC Data Center for Astrophysics of the University of Geneva
	 chemin d'\'Ecogia, 16 1290 Versoix Switzerland\\
	\email{carlo.ferrigno@unige.ch}
         \and
          International Space Science Institute (ISSI) Hallerstrasse 6, CH-3012 Bern, Switzerland
          \and
          INAF - Osservatorio Astronomico di Roma, Via Frascati 33, I-00044 Rome, Italy
          \and
          INAF - Osservatorio Astronomico di Brera, via Emilio Bianchi 46, I-23807 Merate (LC), Italy
          \and
          ESA, European Space Astronomy Centre (ESAC), PO Box 78, E-28691 Villanueva de la Ca\~nada (Madrid), Spain
          \and
          Universit\'a degli Studi di Cagliari, Dipartimento di Fisica, SP Monserrato-Sestu, KM 0.7, I-09042 Monserrato (CA), Italy
          \and 
          INAFâ Osservatorio Astronomico di Cagliari, Poggio dei Pini, Strada 54, I-09012 Capoterra (CA), Italy
         }

\date{Received ---; accepted ---}

\abstract
	{The burst-only source \src\ was discovered in a high X-ray-active state, while during an \textsl{INTEGRAL} 
	observations of the Galactic bulge on 2010 April 10. Pulsations at 518\,Hz were discovered in the \textsl{RXTE} data, 
	confirming previous suggestions of possible associations between burst-only sources and accreting millisecond 
	X-ray pulsars. The subsequent discovery of X-ray eclipses made
	Swift\,J1749.4$-$2807 the first eclipsing accreting millisecond X-ray pulsar.} 
	{We obtain additional information on \src\ and other burst-only sources. }
	{We report on the results of a monitoring campaign on the source, carried out for about two weeks with the 
	\swift,\ \inte,\ and \rxte\ satellites.} 
	{The observations showed that the X-ray spectrum (energy range 0.5--40\,keV) of \src\ during the entire event 
	was accurately modeled by an absorbed power-law model ($N_{\rm H}$$\simeq$3$\times$10$^{22}$~cm$^{-2}$, 
	$\Gamma$$\simeq$1.7).  X-ray eclipses were 
	also detected in the \swift\ data and provides a clear evidence of a dust-scattering halo 
	located along the line of sight to the source. 
	Only one type-I X-ray burst was observed throughout the two-weeks long monitoring.    
	The X-ray flux of \src\ decayed below the detection threshold of \swift\,/XRT about 11~days after the discovery, 
	in a exponential fashion (e-folding time of $\tau$=12$^{+7}_{-3}$~days).} 
	{We compare the properties of the outburst observed from \src\ with those of the previously known millisecond X-ray 
	pulsars and other transient low mass X-ray binaries.} 
	

\maketitle

\section{Introduction}
\label{sec:intro}

Low mass X-ray binaries (LMXBs) consist of a low-mass donor star ($<$1~M$_{\odot}$) 
and a compact object that accretes matter through an accretion disk. 
Most LMXBs are transients, i.e. they undergo week-to-month long 
outbursts during which the accretion takes place at high rates, giving rise to 
typical (``persistent'') X-ray luminosity of $\sim$10$^{36-38}$~erg/s. This is   
$\gtrsim$100 times higher than the X-ray luminosity displayed in quiescence \citep[see e.g.]
[for reviews]{bildsten98,campana98,liu07}. Depending on the peak X-ray luminosity 
reached during the outburst ($L_{\rm peak}$), transient LMXBs are historically classified as 
``bright transient'' ($L_{\rm peak}$$\simeq$10$^{37-38}$~erg/s), ``faint transient'' 
($L_{\rm peak}$$\simeq$10$^{36-37}$~erg/s), or ``very faint transients'' 
\citep[$L_{\rm peak}$$\leq$10$^{36}$~erg/s, see e.g.,][]{wijnands06,campana09}. 
The outbursts of transient LMXBs are usually interpreted in terms of disk instability models  
\citep[see e.g.,][]{frank02}. In a number of these sources,  
an accreting neutron star (NS) as a compact object has been unambiguously 
identified by the detection of type-I X-ray bursts and/or coherent pulsations. 
The former are bright ($\sim$10$^{38}$~erg/s) and 
short ($\sim$10-60~s) flares   
that result from thermonuclear explosions occurring 
in the material accreted at the NS surface \citep[see e.g. ][for reviews]{lewin93,strohmayer06}.   

A peculiar subclass of transient LMXBs was discovered about 10 years ago, mainly thanks to 
the long-term monitoring of the Galactic center carried out with the \beppo\ wide field cameras   
\citep[WFC,][]{heise99,cocchi01,cornelisse02,cornelisse04}. 
At odds with the previously known transient LMXBs, 
these systems appeared to be characterized by 
no persistent emission before and after the occurrence 
of a type-I X-ray burst \citep{cornelisse02}.  
The upper-limits inferred for their persistent emission were significantly lower than  
the persistent emission of the bright transients (10$^{35-36}$~erg/s) but not very tight 
due to the modest sensitivity of the \beppo\,-WFCs. For this reason, 
they were collectively termed ``burst-only sources''. 
Only in a few cases, a few ks-long \chan\ observations carried out after the discovery of a  
type-I X-ray burst from some of these sources, were able to reveal the presence of a weak persistent emission at a position 
consistent with that of the X-ray burst. Typical (quiescent) luminosities derived from 
these detections were $\sim$10$^{32}$-10$^{33}$~erg/s \citep{cornelisse02b}.   

Owing to their peculiar behavior and the discovery of several new members of this 
class \citep[see e.g.,][]{chelovekov07,degenaar09}, the burst-only sources 
have attracted increasing interest in the past few years \citep[see also][]{delsanto10}. 
Systems undergoing type-I X-ray bursts when subject to 
very low accretion rates allow us to test models of thermonuclear burning 
in a regime that is still poorly explored \citep[see e.g.,][]{cooper07,peng07}. 
Moreover, the low persistent luminosity in quiescence and the spectral 
properties during the outbursts 
led to the suggestion that the burst-only sources could be linked 
to very faint transients and accreting 
millisecond X-ray pulsars \citep[AMSPs,][]{wijnands06,campana09,trap09}.  

In this paper, we report on the recent discovery of an intense X-ray 
activity from the burst-only source \src.\ 
We studied the timing and spectral properties of the 
source during this event by exploiting all the available 
\swift,\ \inte,\ and \rxte\ data, and  
report on the detection of millisecond pulsations in the X-ray 
emission of this source. 
The results presented here provide a strong support in favor of the association 
between burst-only sources and AMSPs. 
In Sect.~\ref{sec:source}, we describe previous observations of \src,\ 
and in Sect.~\ref{sec:results} our data 
analysis and results. Our discussion 
and conclusions are presented in Sect.~\ref{sec:discussion}.

\subsection{SWIFT~J1749.4$-$2807}
\label{sec:source}

\begin{table*}
\scriptsize
\caption{Observation log of \src.\ EXP indicates the exposure time in ks of each observation 
(for \inte, it is given separately for ISGRI and JEM-X2 in parentheses), 
$N_{\rm H}$ the absorption column density in units of 10$^{22}$~cm$^{-2}$, and $\Gamma$ the power-law 
photon index of the X-ray spectrum. For \inte, an exponential cutoff energy was introduced in the 
fit (energy fixed at 20~keV). $F_{\rm obs}^{c}$ is the observed flux and is given in the 
0.5-10\,keV energy band for \swift\,/XRT, and in the 3-20~keV energy band for the 
\rxte\,/PCA and \inte\,/ISGRI+JEMX-2 in units of $10^{-10}\,\mathrm{erg/s/cm^{2}}$.}
\begin{tabular}{@{}llllllll@{}}
\hline
\noalign{\smallskip}
\swift\ & & & &\\
\noalign{\smallskip}
\hline
\noalign{\smallskip}
OBS ID & START TIME & STOP TIME & EXP & $N_{\rm H}$ & $\Gamma$ & $F_{\rm obs}$ & $\chi^2_{\rm red}$/d.o.f. \\
\noalign{\smallskip}
\hline
\noalign{\smallskip}
00031686001 &  2010-04-11 23:39:16 & 2010-04-12 04:48:49 &  4.9  & 3.2$\pm$0.3  &  2.6$\pm$0.2   & 3.0$\pm$0.2&  0.97/140 \\
\noalign{\smallskip}
00031686002 &  2010-04-12 23:54:41 & 2010-04-14 01:32:48 &  2.6  &  3.0$\pm$0.4  &  2.3$\pm$0.2 & 2.6$\pm$0.3  &  0.99/85 \\ 
\noalign{\smallskip}
00031686002$^a$  &  2010-04-12 23:54:35 & 2010-04-14 01:31:11 &  0.07  &  4.1$^{+2.6}_{-1.7}$  &  2.4$^{+1.0}_{-0.8}$ &  2.3$^{+0.3}_{-2.3}$  &  (36.6/39)$^c$ \\
\noalign{\smallskip}
00031686003 &  2010-04-16 16:07:00 & 2010-04-16 18:10:00 &  3.1 & 2.9$\pm$0.3  &  2.2$\pm$0.2  & 1.3$\pm$0.1  &  0.90/87 \\
\noalign{\smallskip}
00031686004 &  2010-04-17 11:08:00 & 2010-04-17 13:10:00 &  2.9 & 2.7$\pm$0.5  &  1.9$\pm$0.2 & 9.0$\pm$1.0  & 0.99/87  \\ 
\noalign{\smallskip}
00031686005 &  2010-04-18 03:34:00 & 2010-04-18 07:07:00 &  3.0 & 2.7$\pm$0.4  &  1.8$\pm$0.2  & 5.7$\pm$0.4  &  0.85/78  \\
\noalign{\smallskip}
00031686006 &  2010-04-20 16:29:00 & 2010-04-20 23:04:00 &  1.1 & 1.0$^{+3.0}_{-1.0}$  &  1.0$^{+1.1}_{-0.8}$   & $(1.0^{+0.3}_{-1.0})\times 10^{-1}$  & (15.0/13)$^c$  \\
\noalign{\smallskip}
00031686007 &  2010-04-21 16:19:00 & 2010-04-21 18:26:00 &  1.6 & 2.7 (fixed)  &  1.8 (fixed)   & $(6.5^{+2.3}_{-2.3})\times 10^{-3}$ & -    \\
\noalign{\smallskip}
00031686008$^{b}$ &  2010-04-23 18:25:00 & 2010-04-23 23:38:00 &  3.3 & 2.7 (fixed)  & 1.8 (fixed)   & $<3.3\times 10^{-3}$ & -  \\
\noalign{\smallskip}
00031686009$^{b}$ &  2010-04-24 20:23:00 & 2010-04-24 23:21:00 &  2.2 & 2.7 (fixed)  & 1.8 (fixed)  &  $<6.5\times 10^{-3}$ & -   \\ 
\noalign{\smallskip}
00031686010$^{b}$ &  2010-04-25 21:32:00 & 2010-04-25 23:39:00 &  1.0 & 2.7 (fixed)  & 1.8 (fixed)  &  $<8.8\times 10^{-3}$ & -   \\  
\noalign{\smallskip}
\hline
\noalign{\smallskip}
\rxte\ & & & &\\
\noalign{\smallskip}
\hline
\noalign{\smallskip} 
95085-09-01-00 & 2010-04-14 20:59:12 & 2010-04-14 21:34:40 & 1.5 & $3.0 \pm 1.0$ & $1.8 \pm 0.1$ & $2.54_{-0.06}^{+0.03} $ & 1.2/43 \\
\noalign{\smallskip}
95085-09-01-01 & 2010-04-15 16:49:20 & 2010-04-15 17:42:40 & 2.1 & $3.2 \pm 1.0$ & $1.8 \pm 0.1$ & $2.21_{-0.05}^{+0.03} $ & 1.3/43 \\
\noalign{\smallskip}  
95085-09-01-02 & 2010-04-15 21:31:12 & 2010-04-16 00:01:04 & 5.7 & $3.6 \pm 0.6$ & $1.72 \pm 0.04$ & $2.13_{-0.03}^{+0.02} $ & 1.2/43 \\
\noalign{\smallskip}
95085-09-02-00 & 2010-04-16 14:46:24 & 2010-04-16 15:41:52 & 2.5 & $3.1 \pm 1.4$ & $1.7 \pm 0.1$ & $2.15_{-0.08}^{+0.04} $ & 1.3/43 \\
\noalign{\smallskip}
95085-09-02-01 & 2010-04-16 16:32:16 & 2010-04-16 17:33:36 & 2.2 & $2.9 \pm 1.5$ & $1.7 \pm 0.1$ & $2.06_{-0.09}^{+0.04} $ & 1.3/43 \\
\noalign{\smallskip}
95085-09-02-02 & 2010-04-16 22:37:20 & 2010-04-17 00:24:32 & 2.6 & $3.4 \pm 0.4 $ & $1.7 \pm 0.1$ & $2.07_{-0.09}^{+0.04} $ & 1.0/43 \\ 
\noalign{\smallskip}
95085-09-02-09 & 2010-04-17 19:20:16 & 2010-04-17 23:21:36 & 8.0 & $3.6 \pm 1.6 $ & $1.7 \pm 0.1$ & $1.23_{-0.06}^{+0.03} $ & 1.4/43 \\
\noalign{\smallskip}
95085-09-02-03 & 2010-04-18 15:25:20 & 2010-04-18 17:04:32 & 2.7 & 3 (fixed) & $1.6 \pm 0.1$ & $0.96_{-0.03}^{+0.03} $ & 0.9/44 \\
\noalign{\smallskip}
95085-09-02-06 & 2010-04-19T07:18:24 & 2010-04-19T07:32:00 & 0.7 & 3 (fixed) & $1.6 \pm 0.2$ & $0.71_{-0.09}^{+0.06}$ & 1.5/44 \\
\noalign{\smallskip}
95085-09-02-04 & 2010-04-19 19:31:12 & 2010-04-19 22:24:32 & 6.4 & 3 (fixed) & $1.6 \pm 0.1$ & $0.51_{-0.04}^{+0.03} $ & 1.3/44 \\
\noalign{\smallskip}
95085-09-02-07 & 2010-04-20T08:12:16 & 2010-04-20 08:42:56 & 1.0 & 3 (fixed) & $1.5 \pm 0.2$ & $0.40_{-0.11}^{+0.04} $ & 1.0/44 \\
\noalign{\smallskip}
95085-09-02-05 & 2010-04-20 16:01:20 & 2010-04-20 19:15:28 & 3.2 & 3 (fixed) & $1.7 \pm 0.2$ & $0.27_{-0.07}^{+0.02} $ & 1.0/46 \\ 
\noalign{\smallskip}
95085-09-02-11 & 2010-04-20 23:10:24 & 2010-04-20 23:28:32 & 0.8 & 3 (fixed) & 1.7 (fixed) & $<1.3\times 10^{-1}$ & - \\
95085-09-02-08$^d$ & 2010-04-21 13:58:24 & 2010-04-21 14:55:28 & 2.4 & - & - & - & - \\
95085-09-02-10$^d$ & 2010-04-21 15:29:20 & 2010-04-21 17:16:32 & 3.2 & - & - & - & - \\
\noalign{\smallskip}
\hline
\noalign{\smallskip}
\inte\ & & & &\\
\noalign{\smallskip}
\hline
\noalign{\smallskip} 
0720001/0018$^e$ & 2010-04-07 13:31:59 & 2010-04-07 17:13:34 & 15(9) & - & - &$<3$ & \\
\noalign{\smallskip} 
0720001/0019 & 2010-04-10 22:00:00 & 2010-04-11 01:41:36 & 8.4(6.5) & 3 (fixed) & $1.3\pm0.3$  & $3.3_{-1.3}^{+0.6}$ & 0.5/16 \\
\noalign{\smallskip} 
0720001/0020 & 2010-04-13 15:00:00 & 2010-04-13 18:41:45 & 8.9(11.3) & 3 (fixed) & $1.6\pm0.4$  & $2.8_{-1.1}^{+0.6}$  & 0.3/15 \\
\noalign{\smallskip} 
0720001/0021$^e$ & 2010-04-21 20:24:43 & 2010-04-22 00:06:28 & 6.9(6.7) & - & - & $<4$ & \\
\noalign{\smallskip}
\hline
\noalign{\smallskip}
\multicolumn{8}{l}{$^{a}$: This \swift\,/XRT observation was carried out in WT mode. $^{b}$: 3$\sigma$ upper limit. $^{c}$: In these cases this value is C-stat/d.o.f.} \\
\multicolumn{8}{l}{$^{d}$: Observations used for the Galactic diffuse emission estimation.}\\
\multicolumn{8}{l}{$^{e}$: These are $3\sigma$ upper limits on the source X-ray flux (3--20~keV) estimated from the JEM-X2 mosaics.}\\
\end{tabular}
\label{tab:log}
\end{table*} 

\src\ was discovered in 2006 during a bright type-I X-ray burst that was initially recorded as a potential   
gamma ray burst \citep[GRB060602B,][]{wijnands09}. \swift\,/XRT started to follow-up the evolution of the source  
from 83\,s after the BAT trigger, and monitored the source outburst for the following 8~d. The source X-ray flux 
decayed in an approximately power-law fashion (with index $\sim$-1), fading 
below the detection threshold of \swift\,/XRT in less than $10^6$\,s. 
The BAT spectrum extracted at the peak of the type-I X-ray burst could be described by a black-body (BB) model 
(kT=2.9$^{+0.4}_{-0.3}$~keV) and provided an upper limit to the source 
distance of $6.7\pm1.3$\,kpc \citep[by assuming that the peak X-ray luminosity of the burst 
corresponded to the Eddington value, uncertainties at 90\% c.l.;][]{wijnands09}.  
During the outburst, the XRT spectrum could be described well by an 
absorbed power-law model. The measured column density was consistent with being 
constant throughout the event at a value 
$N_{\rm H}$=4$\times$10$^{22}$~cm$^{-2}$, while the photon index 
was observed to decrease from $2.7^{+1.5}_{-1.1}$ to 
$0.5\pm1.3$ \citep{campana09}. 
The estimated X-ray luminosity was $\sim$10$^{36}$~erg/s at the peak ot the outburst and 
$\sim$10$^{32}$~erg/s in the latest \swift\ observation available. 

\citet{wijnands09} also reported on three serendipitous detections of \src\ 
in archival \XMM\ observations. The first observation was carried out 
on 2000 September 23, and the second two were performed on 2006 September 22 and 26. 
The count-rate of the source in the three \XMM\ observations was too low 
to extract any meaningful spectral information. By assuming an absorbed power-law model with $\Gamma$=2 
and $N_{\rm H}$=3$\times$10$^{22}$~cm$^{-2}$, \citet{wijnands09} estimated a 2-10~keV 
unabsorbed flux of $\sim$(1-2)$\times$10$^{-13}$~erg/cm$^{2}$/s. This corresponds to a luminosity of 
(3-6)$\times$10$^{32}$~erg/s (assuming a distance to the source of 7~kpc) and is compatible, to 
within the errors, with that measured about 6~days after the outburst discovered with \swift.\  

\src\ was detected again in a high X-ray luminosity state on 2010 April 10, during 
the \textsl{INTEGRAL} Galactic bulge monitoring program \citep{pavan10,kuulkers07}. 
\swift\ and RXTE target of opportunity observations (ToO) were immediately requested, and these monitored the source 
outburst for about two weeks. \rxte\ observations detected coherent pulsations in the X-ray emission of the source 
at 518\,Hz and its second harmonic \citep{altamirano2010a,bozzo2010}. From these data, a preliminary orbital 
solution was first obtained by \citet{belloni10}, and then refined by \citet{strohmayer2010} by using a pulse 
phase-coherent technique.  
The latter authors derived an orbital period of $31740.345\pm 0.04$\,s (8.8168 hr), a projected semi-major 
axis of $a \sin(i) = 1899.53 \pm 0.01$\,lt-ms, and a barycentric frequency for the second harmonic  
(i.e., twice the spin frequency) of $1035.840025$\,Hz $\pm0.4\,\mu$Hz. The time of the ascending node was 
$2455301.1522672 \pm0.0000014$\,JD (TDB). This solution implied a mass function 
of 0.05463$\pm$0.00018\,$M_{\odot}$ and a minimum mass for the companion of 0.475~$M_{\odot}$
(assuming a NS of 1.4$M_{\odot}$). The discovery of an X-ray eclipse in the \rxte\ light curve  
was reported by \citet{markwardt10a}; this is the first eclipse observed from an accreting millisecond 
X-ray pulsar. The most accurate source position to date was provided by \citet{yang10} at 
$\alpha_{\rm J2000}$=17$^{\rm h}$49$^{\rm m}$31$\fs$80 and 
$\delta_{\rm J2000}$=-28${\degr}$08$\arcmin$04$\farcs$9, with a 90\% confinement radius of 1.9$\farcs$,
based on \swift\,/XRT observations.   

\section{Data analysis and results}
\label{sec:results} 

\subsection{\inte\ data}

\inte\ data were analyzed using the OSA software (v.9) released by the ISDC \citep{isdc}. 
We considered data from both IBIS/ISGRI \citep{lebrun03} and JEM-X2 \citep{lund03} in the 20--40\,keV and 
3--23\,keV energy ranges, respectively.  
The average fluxes and spectra for JEM-X2 were extracted from 
the mosaic images, as recommended in the case of weak 
sources\footnote{See also http://isdcul3.unige.ch/Soft/download/osa/ osa\_doc/osa\_doc-9.0/osa\_um\_jemx-9.1.pdf}. 
The ISGRI spectrum was extracted using standard procedures. The detail of the \inte\ observations is given in Table~\ref{tab:log}. 
The source was not detected during the observations carried out on 2010 April 7 and 21-22. 
\src\ was clearly detected (8$\sigma$) during the two observations carried out from 2010 April 10 to 13. 
The simultaneous ISGRI+JEM-X2 spectra could be reasonably well described by  
a cutoff power-law model of photon index 1.3-1.6 (we fixed the cutoff energy at 20 keV and the absorption column density at 
3$\times$10$^{22}$~cm$^{-2}$, see below); the estimated flux was 3$\times$10$^{-10}$~erg/s/cm$^{2}$ (3-20~keV).   
For these observations, we also extracted an event list from the entire JEM-X2 detector in the 3--20\,keV band
to search for type-I X-ray bursts. Only one burst was detected: the increase in the source flux at the time of the burst 
is also confirmed by the higher significance of the source detection in the JEM-X2 image extracted during the duration of the burst 
\citep[see also][]{chenevez10}. 
Our analysis of the burst parameters is described below. A search for type-I X-ray bursts was also performed 
in the ISGRI data (we used a list of events selected in the 18-40 keV energy band and with a pixel illumination fraction 
threshold $>$0.75). No statistically significant bursts were detected in these data.

\subsubsection{The type-I X-ray burst}

\begin{figure}
\centering
\includegraphics[scale=0.32]{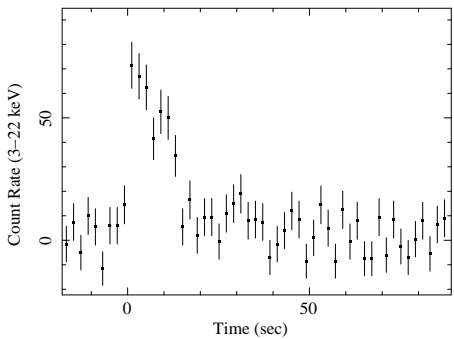}
\caption{The type-I X-ray burst detected by INTEGRAL/JEM-X2 from SWIFT~J1749.4-2807.
The JEM-X (3-20 keV) net light curve is shown (background subtracted). The time bin is 2 s
and the start time of the burst is 2010 April 13 at 16:51:18 (UTC at satellite location).}   
\label{fig:burst} 
\end{figure}

In the \inte\ observation carried out on 2010 April 13, a type-I X-ray burst was detected by JEM-X2.  
The light curve of the burst in the 3-20 keV band and with a time resolution of 2~s is shown in Fig.~\ref{fig:burst}.  
The rise time of the burst was $\sim$1~s, and the start time\footnote{We defined the start time of the burst as the time 
at which the intensity of the source was 10\% of the peak intensity above the persistent level.} was 2010 April 13 
at 16:51:18 (UTC at the satellite location; these values were 
determined by using the JEM-X2 source event list rebinned to have a time resolution of 0.5~s).   
The relevant burst parameters are reported  in Table~\ref{tab:burst}.
Owing to the relatively low signal-to-noise ratio (S/N), a time-resolved spectral analysis of the burst could not be carried out,
thus no signature of a possible photospheric radius expansion \citep[PRE; see e.g.,][]{lewin93} could be identified. 
We determined the flux at the peak of the burst by fitting the spectrum of the initial 4~s with a BB model (the $N_{\rm H}$ was fixed at 
$3\times10^{22}$~cm$^{-2}$, see Table~\ref{tab:log}). 
The persistent spectrum extracted from a time interval close to the burst was used as a background in the fit. 
The best-fit BB temperature and radius at the peak were $kT_{\rm bb,peak}= 2.3^{+0.7}_{-0.5}$ keV 
and $R_{\rm bb, peak}= 4.8^{+1.2}_{-0.3}$~km, respectively 
(for  a source distance of 7 kpc) with a $\chi^{2}_{\rm red}/d.o.f.=0.8/3$. 
In Table ~\ref{tab:burst}, we report the decay time, $\tau_\mathrm{lc}$, of the burst as measured 
by fitting the observed light curve with an exponential function; this is to be compared with $\tau$, 
the burst duration that is obtained by dividing the burst fluence by the peak flux 
\citep[see e.g.][]{lewin93}.  
We note that the two values are in agreement to within the errors. 
Additional comments on the type-I X-ray burst are given in Sect.~\ref{sec:discussion}. 

\begin{table}
\centering
\scriptsize
\caption{The type-I X-ray burst parameters.}
\begin{tabular}{@{}lc@{}}
\hline
\noalign{\smallskip}
Start time$^a$ & 55\,299.70229 \\
$\tau_\mathrm{lc}$ (s)$^b$    & $11.3\pm2.1$ \\
$F_{\rm peak}$ (10$^{-8}$~erg/cm$^{2}$/s)$^c$  &  $3.0\pm0.6$\\
$f_b$  (10$^{-7}$ erg/cm$^{2}$)$^d$ & $3.4\pm0.2$\\
$F_{\rm pers}$ (10$^{-10}$~erg/cm$^{2}$/s)$^e$ & $8.0\pm4.0$   \\
$\gamma  \equiv F_{\rm pers}/F_{\rm peak}$  (10$^{-2}$) &  $2.7\pm1.9$\\
$\tau$ (s) $\equiv f_b / F_{\rm peak} $  &  $10.4\pm2.5$\\
\noalign{\smallskip}
\hline
\noalign{\smallskip}
\multicolumn{2}{l}{$^{a}$ MJD, at the satellite reference frame.}\\
\multicolumn{2}{l}{$^{b}$ Burst e-folding decay time. $^{c}$Net unabsorbed peak flux.} \\ 
\multicolumn{2}{l}{$^{d}$ Net unabsorbed burst fluence.} \\ 
\multicolumn{2}{l}{$^{e}$ Unabsorbed persistent flux (0.1--100 keV).}\\ 
\end{tabular}
\label{tab:burst}
\end{table} 

\subsection{\rxte\ data}
\label{sec:rxte} 

\rxte\,/PCA \citep{jahoda96} data analysis and spectral extraction were carried out with the 
standard tools available in {\sc HEASOFT} (V 6.7). 
Since the source was relatively faint for the instruments on-board \rxte\ and is 
located in the Galactic bulge, we paid particular attention to estimate the background as accurately as possible. 
For this purpose, we excluded from the spectral analysis, performed using the \texttt{standard2} mode, 
the data from the PCU0 detector, because the propane veto layer  
stopped working in May 2000, leading to significantly worse instrumental noise subtraction. 
Moreover, we considered only data from the upper anode layer to reduce the systematic error in the instrumental noise.
 
The large contribution from the Galactic diffuse emission to the \textsl{RXTE/PCA} X-ray fluxes, poses a serious problem 
for data processing. We used the latest observations, performed after 2010 April 20 to estimate 
the background for the utilized PCU configurations (Table~\ref{tab:log}).  After 2010 April 20, the source 
flux measured by \swift\,/XRT (see Table~\ref{tab:log}) decreased to below 
10$^{-11}$~erg/cm$^{2}$/s, indicating that the count-rate in the last two \rxte\ observations was dominated by the Galactic 
ridge emission \citep{valinia98}. We did not use the HEXTE data, since both units stopped rocking on 2009 December 14 
and the background estimation cannot be performed satisfactorily with 
the available software at the time of writing (the application of the method used for PCA 
does not give reliable results because of the instrumental variable background induced 
by high energy particles). 

A log of the \rxte\ observations and a summary of the results 
of the analysis of these data is provided in Table~\ref{tab:log}. 
Here, all the spectra were fit within {\sc xspec} in the 3-23~keV energy range with an absorbed 
power-law model, which always provided satisfactory fits. We checked that the relatively large values of the 
$\chi^2$ were due to residuals in the background subtraction 
and not to the presence of additional spectral components.  

\begin{figure}
\centering
\includegraphics[scale=0.4]{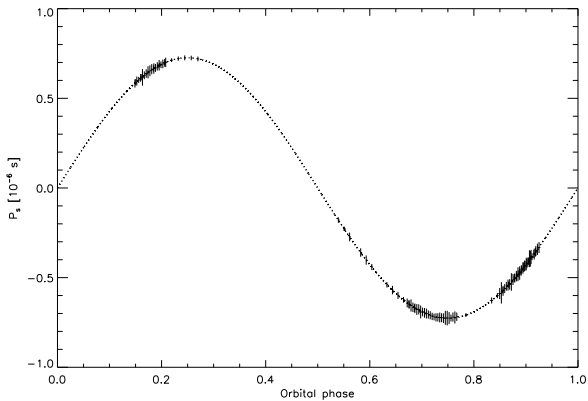}
\caption{Measurements of the spin period of \src\ in the \rxte\ data. 
The spin period is computed in each case by using a time window with a duration of between 
200~s and 1000~s (depending on the source intensity) and fitting with a Gaussian function 
the peak in the $Z^2$-statistics corresponding to the first overtone of the pulse. 
From these measurements, we derived the orbital parameters reported 
in Table~\ref{tab:orbitalsolution}. The corresponding orbital solution is represented in this figure 
with a dotted line.}   
\label{fig:folded} 
\end{figure}

For the timing analysis, we used the \rxte\ event data (mode \texttt{E\_125us\_64M\_0\_1s}) with 64 energy channels 
and 125\,$\mu s$ time resolution. To maximize the S/N, we selected the events in all the active
PCUs and layers. The barycenter correction was applied using the tool {\sc faxbary} 
and the most precise available source position to date (see Sect.~\ref{sec:source}).   
Pulsations at the spin frequency of the NS and its second harmonic were clearly detected 
in all the observations by using the $Z^2$-statistics  \citep{buccheri1983,markwardt2002}, in agreement 
with the results reported by \citet{altamirano2010a} and \citet{bozzo2010}.  
An orbital solution for \src\ was obtained from the timing analysis based on the frequency modulation of the 
signal (see Fig.~\ref{fig:folded} and Table~\ref{tab:orbitalsolution}). It gave results consistent 
with those reported by \citet{strohmayer2010}, though with larger uncertainties,
since they derived a phase-coherent solution. 
In the following we use their value for the orbital period, projected semi-major axis, barycentric pulse 
frequency, and epoch of the ascending node. 
\begin{table}
\centering
\tiny
\caption{Parameters of the orbital solution of \src,\ as derived from the \rxte\ observations 
(errors are given at 1$\sigma$ c.l. on the last digit).}
\begin{tabular}{@{}lc@{}}
\hline
\noalign{\smallskip}
$P_{\rm spin}$~(ms) & 1.9308000(2)\\ 
$a$$sin(i)$~(lt-s) & 1.8999(8)  \\
$P_{\rm orb}$~(s) & 31\,740.7(8) \\
$T_{\rm mid eclipse}$~(MJD) & 55\,300.37672(9) \\ 
$\delta$$T_{\rm eclipse}$~(s)$^a$ & 2190(4)  \\ 
$f(m)$~($M_{\odot}$) & 0.05456(7) \\
\noalign{\smallskip}
\hline
\noalign{\smallskip} 
\multicolumn{2}{l}{$^{a}$: This correspond to the best determined eclipse} \\ 
\multicolumn{2}{l}{duration from the \rxte\ observations (see Sect.~\ref{sec:rxte}).} \\ 
\end{tabular}
\label{tab:orbitalsolution}
\end{table} 

\begin{figure}
\centering
\includegraphics[scale=0.55,angle=0]{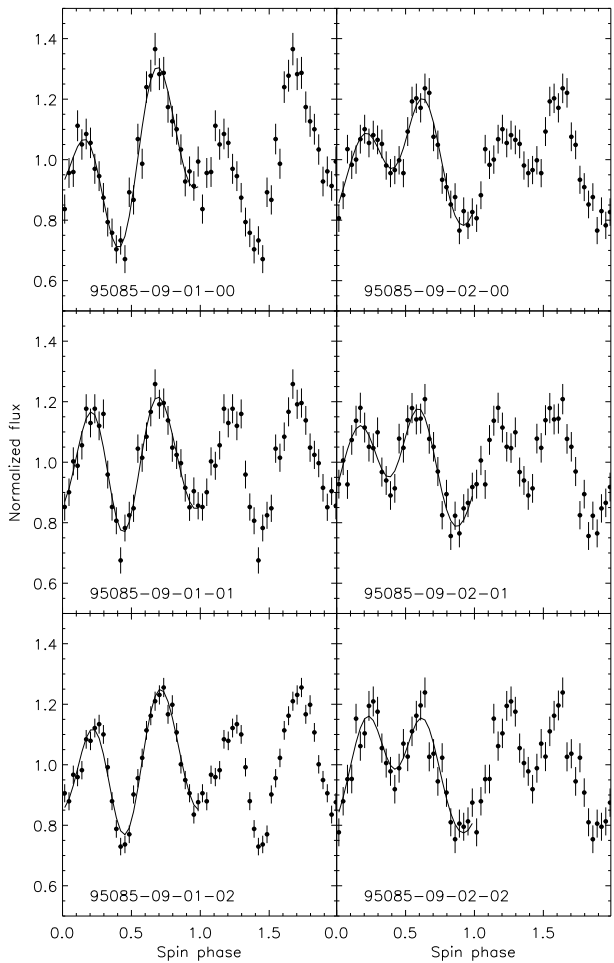}
\caption{Pulse profiles of \src\ in the \rxte\ observations (the observation ID. is indicated in each figure) 
normalized for the corresponding average count-rate and folded with respect to the same reference time 
(53\,888.9854166670 MJD).  The energy band is 3--23~keV (data from PCU2).  In the observation 95085-09-02-02, the eclipse was excluded. The solid line in each figure is obtained by truncating the Fourier series expansion of the pulses 
profiles to the first three terms.} 
\label{fig:timepulses}
\end{figure}

To study the variations in the amplitude of the fundamental and second harmonic throughout the 
\rxte\ monitoring \citep{bozzo2010}, we used the value of the source spin period reported by \citet{strohmayer2010} and folded the light curves  
into time intervals of 100\,s to produce in each interval a pulse profile, $P_t(\phi)$, in 32 phase bins.
We rebinned these pulses adaptively by choosing appropriate time intervals of durations $t_{\rm stop}$-$t_{\rm start}$ 
that permitted us to obtain a S/N$\gtrsim$50 (we were careful not to combine observations 
separated by more than $\sim$1~day). The characteristic amplitude of the 
first two Fourier components in each rebinned pulse profile was then computed using the equation 
\begin{equation}
A_n(t)=\sqrt{\left(I^n_c(t)\right)^2+\left(I^n_s(t)\right)^2}\,,
\end{equation}
where $n=1,2$ is the Fourier order
\begin{equation}
I^n_c(t) = \int P_t(\phi)  \cos(n\phi) \mathrm{d}\phi\,, 
\end{equation}
\begin{equation}
I^n_s(t) = \int  P_t(\phi)  \sin(n\phi)  \mathrm{d}\phi\,, 
\end{equation}
and $t$=$t_{\rm start}$+($t_{\rm stop}$-$t_{\rm start}$)/2.  
This method has the advantage that the statistical uncertainties in 
$A_{\rm n}$(t) can be straightforwardly computed 
by propagating the errors from the pulse 
profiles. 
\begin{figure}
\centering
\includegraphics[angle=0,scale=0.45]{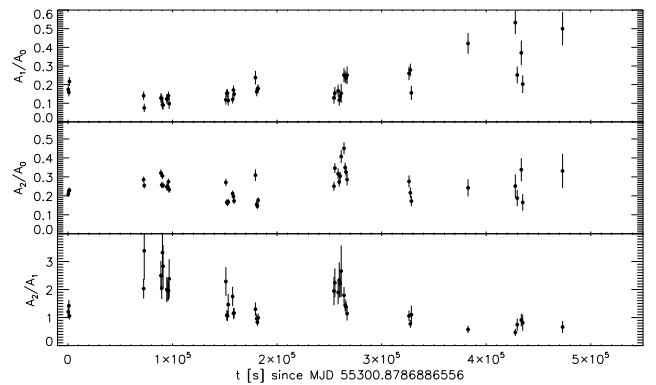}
\caption{Fourier analysis of the pulsed signal throughout the \rxte\ monitoring of 
\src\ : $A_{0,1,2}$ are the powers of the first three Fourier coefficients. The pulse profiles have been rebinned to obtain a S/N at least 50 and only detections at more than $3\sigma$ of $A_{1,2}$ are reported.}
\label{fig:power}
\end{figure} 
From Fig.~\ref{fig:power}, we note that during the observation 95085-09-01-01 ($\sim10^5$~s after the first 
\rxte\ observation) the power of the second harmonic became higher than that of the fundamental, 
in agreement with the  results reported by \citet{bozzo2010}. 
The relative power of the two components, $A_1$/$A_2$, varies with time: the second harmonic was 
clearly dominant during the early stage, before 2010 April 18, (i.e. $t<3\times10^5$\,s in Fig.~\ref{fig:timepulses}), while the fundamental became more prominent in the later stages, 
($t>3\times10^5$\,s in Fig.~\ref{fig:timepulses}). 

To investigate the origin of the behavior of the fundamental and second harmonics with time, we 
studied the source pulse profile in different energy bands. 
Because of the relatively low count rate of the source during the last 
part of the \rxte\ monitoring (see Table~\ref{tab:log}), we used for this analysis only 
the data accumulated prior to 2010 April 17. The pulse profiles of \src\ in each observation were extracted 
in 32 phase bins and corrected for the background by using the technique described above.  
The results are shown in Fig.~\ref{fig:timepulses}, 
where the pulses are phase connected, since the folding reference time is the same. 
The double peak structure of the pulse profile visible in this figure 
explains the higher power in the second harmonic with respect to the fundamental spin frequency, and 
naturally suggests that we are observing at the same time the emission from the 
two polar caps of the NS with roughly the same intensity. 
Furthermore, a visual inspection of the pulses extracted in each observation   
reveals a significant change in the pulse profile with time. 
In particular, when the second harmonic became dominant (observation 95085-01-01), the two peaks 
look very similar (see Fig.~\ref{fig:timepulses}). 
This might be due to a variation in the accretion flow geometry in the vicinity of the polar caps of the NS, 
or alternatively to the occultation of at least part of one polar cap by the
inner region of the accretion disk (see also Sect.~\ref{sec:discussion}). 
A detailed modelling of the pulse profile and their changes will be presented elsewhere. 

\begin{figure}
\centering
\includegraphics[width=7cm,angle=0]{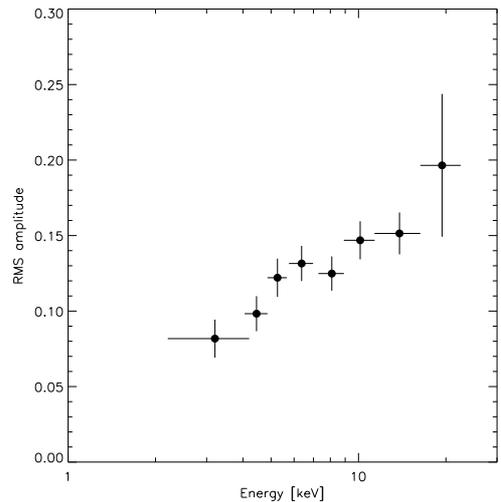}
\caption{Fractional rms as a function of the energy derived from the average pulse profiles 
of the \rxte\ observations (see text for further details).} 
\label{fig:pulsedfraction}
\end{figure}

The dependence of the fractional rms \citep[Eq. (2.10) in ][]{fourier} on energy is shown in 
Fig.~\ref{fig:pulsedfraction}.  In this case, we used the pulse profiles with 32 phase bins obtained by summing up all 
the \rxte\ observations prior to 2010 April 17 (total exposure time $\sim$17\,ks). 
The fractional rms slightly increases with energy, as in the results  
reported for other AMSPs \citep[see e.g.,][]{gierlinski05,patruno09}. 

To search for X-ray eclipses \citep{markwardt10a}, we extracted the source 
light curves of each PCU unit with a time resolution of 1~s. 
These light curves were subtracted for instrumental background 
(we used the background model available for faint sources) and scaled to the effective area of 
the full PCA array\footnote{See also \texttt{http://heasarc.gsfc.nasa.gov/docs/xte/ xhp\_proc\_analysis.html}.}. 
The contribution from the diffuse Galactic emission 
was subtracted by using the data collected in the observations 95085-09-02-08 and 95085-09-02-10 (see above).  

These light curves 
were then barycentered and folded by using the orbital solution 
reported by \citet{strohmayer2010}. We found that an eclipse ingress was visible during observation 
95085-09-02-11, whereas the eclipse egress was clearly present in observations 95085-09-02-02 and 95085-09-02-04. 
No other orbital feature was detected. 
To determine the parameters of the eclipse, we separately fit the light curves 
extracted from the three observations with the rectangular step function \citep[see e.g.,][]{mangano04}:  
\begin{eqnarray}
\lefteqn{\frac{A}{\pi} \left[ \tan^{-1} \right. \left( B \left( 0.5 - C - \phi \right) \right) +} \nonumber \\
& \left. \tan^{-1} \left( B \left( \phi  -0.5 - C\right) \right)\right] +D\,,
\end{eqnarray}
where $\phi$ is the orbital phase, $A$, $B$, and $D$ are constants to be determined from the fit, and  
$C$ is the semi-amplitude of the eclipse in phase units. For large values of $B$  
this expression mimics the shape of the light curve of an eclipse with sharp ingress 
and egress, similar to the one we have in our case (the obscuration 
of a NS of 10\,km moving in a $\sim$9~hr wide orbit around a solar-type star is expected to 
occur in $\Delta t \sim 10^{-6}$\,s). We found that $C=(3.450\pm 0.005)\times10^{-2}$, 
$(3.41\pm 0.03)\times10^{-2}$ for the two egresses, and $C=(3.56\pm 0.13)\times10^{-2}$ for the ingress 
(uncertainties are $1\sigma$ c.l.). The corresponding durations of the eclipses are $2190\pm4$\,s, 
$2163\pm17$\,s and $2260\pm80$\,s, which are all compatible with each other within the errors. The duration of the 
egress is constrained by the first observation to be shorter than 2\,s.

\begin{figure}
\centering
\includegraphics[scale=0.29,angle=0]{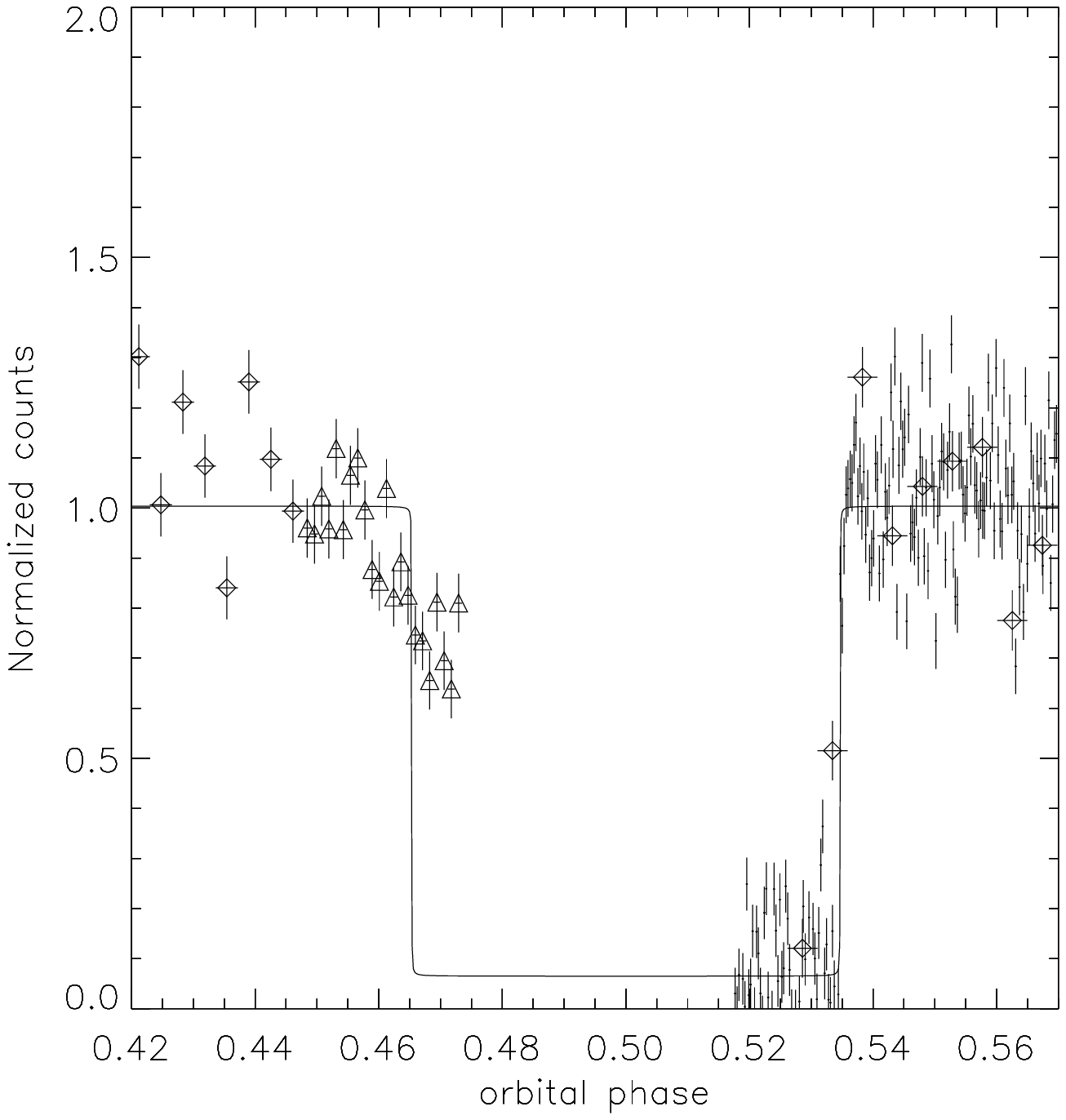}
\includegraphics[scale=0.29,angle=0]{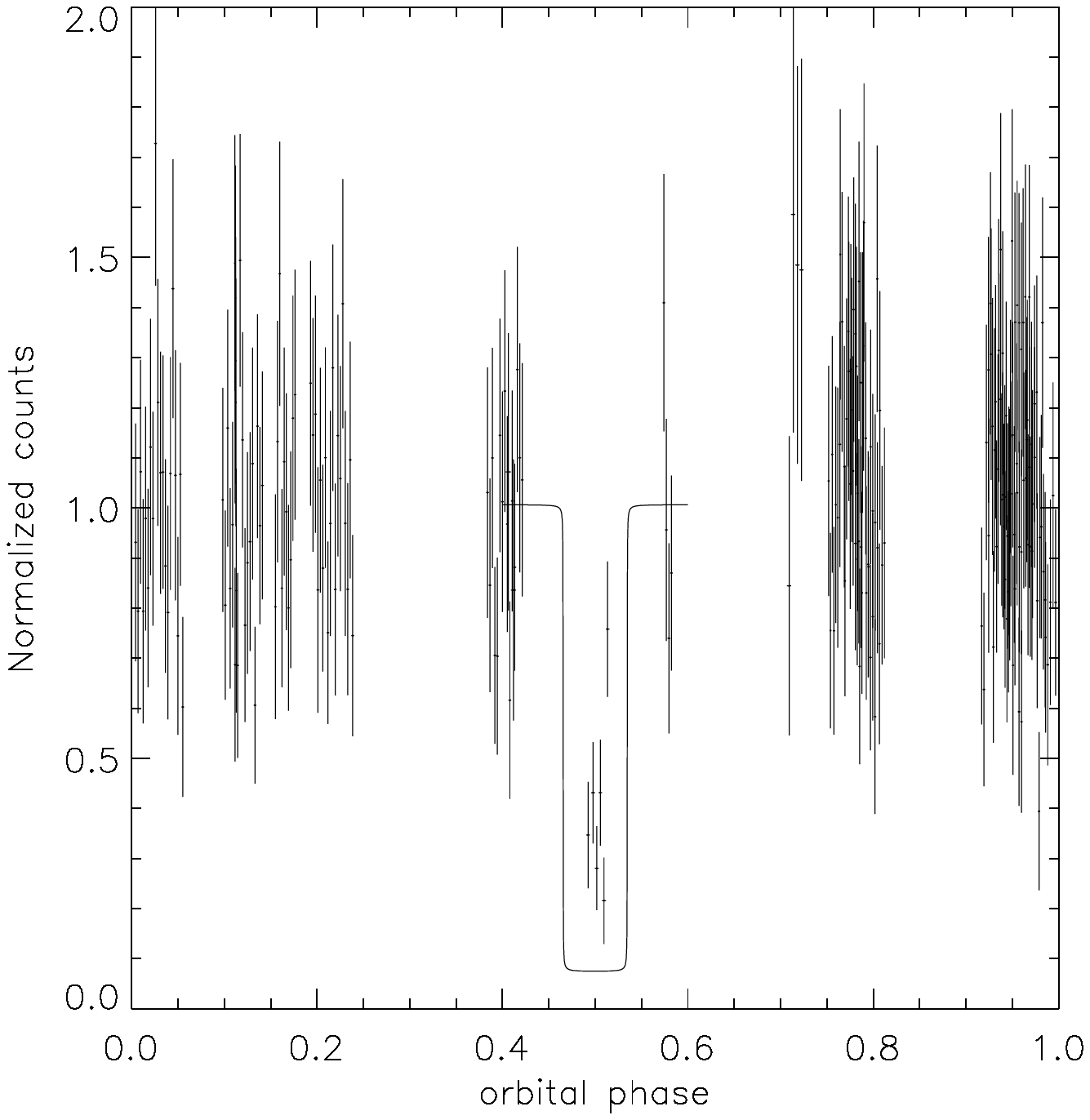}
\caption{{\it Left}: \rxte\ orbital folded light curve of \src.\ Diamonds refer to the observation 95085-09-02-04, 
triangles to 95085-09-02-11, and the other points to the observation 95085-09-02-02. The light curves are normalized to 
the average values outside the eclipse and rebinned to obtain a S/N$\simeq$8 in each time bin. 
{\it Right}: \swift\,/XRT folded light curve of \src\ (0.3-10 keV). The eclipse occurred during the observation 
ID.~00031686002. In both panels, the solid line is the shape of the eclipse determined from the \rxte\ 
observation 95085-09-02-02.}
\label{fig:eclipse}
\end{figure}

The \rxte\ light curve of \src, folded at the orbital period, is shown in Fig.~\ref{fig:eclipse}, 
where we have zoomed around the eclipse. From this figure, we notice that the ingress in the eclipse 
appeared to be less rapid than the egress, and a residual X-ray flux seemed to be present during 
the eclipse. We investigated a possible energy dependence of the eclipse ingress 
profile by extracting source light curves in the 3-5~keV and 6-10~keV energy band and calculating the 
hardness ratio of these light curves\footnote{We defined here the hardness ratio as the ratio 
of the background subtracted \rxte\,/PCA count rate in the hard (6-10~keV) to that in the soft (3-5~keV) energy band versus time.}. 
No evidence of a change in the hardness ratio was found. 
After taking into account the uncertainties in the removal of the background from the \rxte\ observations, 
it is not clear whether the residual X-ray flux visible during the eclipse in Fig.~\ref{fig:eclipse} 
is due to real emission from the source.
The \rxte\ spectrum during the eclipse infers a flux in the 3-20~keV energy band 
of (1.3$\pm$1.1)$\times$10$^{-11}$~erg/cm$^{2}$/s (the power-law photon index was 2.0$^{+1.0}_{-1.5}$).   
To investigate the origin of this X-ray emission, we first searched for pulsations in the X-ray emission 
of \src\ during the eclipse.  We folded the data from the eclipse in observation ID.~95085-09-02-02, which ended 
at MJD 55\,302.96097 TBD, (effective exposure time 530\,s) at the NS spin frequency in 16 phase bins. 
This profile did not show any statistically significant 
modulation. 
To verify this apparent lack of pulsations, we also estimated  
the power of the pulsed emission during the eclipse. We used as a template the pulse profile in the 
observation 95085-02-02 excluding the eclipse (energy band 3-20~keV, fractional rms $(7.3\pm0.6)$\%), 
and computed its cross-correlation coefficient\footnote{The cross-correlation coefficient is defined as $r=\sum_i p_1(i) p_2(i)$, where $p_1$ and $p_2$ are two pulse profiles normalized to have zero average and unitary variance.} with respect to the pulse during the eclipse (exposure 530~s) and during the 
530~s following the eclipse. We measured respectively cross-correlation coefficients of $-0.16\pm0.24$ and $0.72\pm0.12$.

Given the systematic uncertainties in the background subtraction,
providing a clear explanation 
of the residual emission during the eclipse (if any) with \rxte\ can be challenging. 
Much more detailed information can be obtained from the \swift\ observations, 
thus we discuss the origin of the residual emission during the eclipse in Sects.~\ref{sec:swift} 
and  \ref{sec:halo}.  

\subsection{\swift\ data}
\label{sec:swift}

We analyzed the \swift\,/XRT \citep{gehrels04} data collected in both photon counting mode 
(PC) and window timing mode (WT) using standard procedures 
\citep{burrows05} and the latest calibration files available. 
Filtering and screening criteria were applied 
by using {\sc ftools}. 
The barycentric correction was applied to the times of all the event files 
using the online tool {\sc barycorr}. 
We extracted source and background light 
curves and spectra by selecting event grades of 0-2 and 0-12,  
respectively, for the WT and PC mode. Exposure maps were created through 
the {\sc xrtexpomap} task, and we used the latest spectral redistribution 
matrices in the {\sc heasarc} calibration database (v.011). 
Ancillary response files, accounting for different 
extraction regions, vignetting, and point spread function (PSF) corrections, were generated 
by using  the {\sc xrtmkarf} task.  
We corrected the PC data (where required) for pile-up. 
A log of the \swift\ observations is given in Table~\ref{tab:log}. 
For each observation in the table, we extracted the spectrum and derived the  
X-ray flux by fitting an absorbed power-law model.   
Spectra with sufficiently high count rate were rebinned to 
collect at least 20 photons per bin, to permit minimum $\chi^2$ fitting. 
The spectra extracted in observations with ID.~00031686002 (WT mode) and 
00031686006 were characterized by very low count rate; we therefore rebinned these spectra 
to have at least 5 photons per bin and then fit them by using the C-statistics \citep{cash79}.    
During observation ID.~00031686007, the source was clearly detected in the \swift\,/XRT image 
but the exposure was too low to extract a meaningful spectrum.  Therefore, we estimated the source count 
rate of the observation with the {\sc sosta} program (available within the ftool {\sc ximage}), 
and used the resulting count rate within {\sc webpimms}  
to derive the X-ray flux \citep[we assumed the same spectral model of the observation 
ID.~00031686005; see also][]{bozzo09}. 
A similar technique was adopted to estimate a 3$\sigma$ upper limit to the source X-ray flux 
during observations ID.~00031686008,9,10, when \src\ was not detected.   
The results from this analysis are reported in Table~\ref{tab:log}. 
 
The \swift\,/XRT light curves were folded at the orbital period of the source (see Sect.~\ref{sec:rxte})
to search for X-ray eclipses. An eclipse occurred during observation ID.~00031686002.  
A zoom of the light curve of this observation around the eclipse 
is shown in Fig.~\ref{fig:eclipse}. This shows a residual X-ray flux  
during the eclipse (see Sect.~\ref{sec:halo}). 

We also reanalyzed all the observations carried out with \swift\,/XRT during the 2006 outburst 
to search for  X-ray eclipses. 
By using the orbital solution discussed in Sect.~\ref{sec:rxte}, we noticed that some of these observations 
were performed when the source was in eclipse. However, the source count rate was far too 
low to extract a meaningful light curve and spectrum.

\subsection{A scattering halo in the direction of \src\ }
\label{sec:halo} 

\begin{figure}
\centering
\includegraphics[scale=0.13,height=4.2cm]{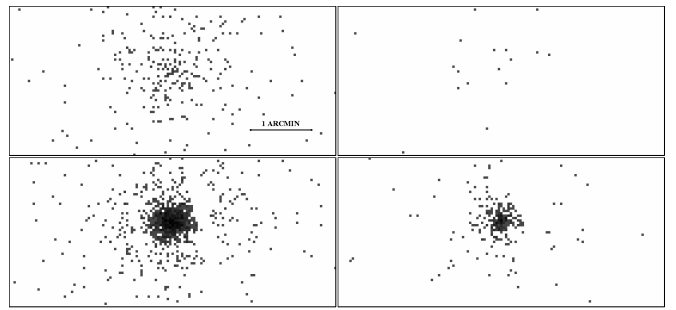}
\caption{{\it Upper panels}: \swift\,/XRT image of \src\  extracted during the eclipse in the observation 
ID.~00031686002 (exposure time 840~s, see Fig.~\ref{fig:eclipse}). The left (right) panel shows the image of the source 
in the 0.3-5~keV (5-10\,keV) energy band. The black line corresponds to a distance of 1~arcmin in the images. 
A 
source is detected at low energies, but not at high energies. 
{\it Lower panels}: the same as for the upper panel, but the images were extracted during the same observation 
outside the eclipse (exposure time 840~s). 
From this image, it is clear that the in-eclipse PSF looks more extended than the point-like PSF observed out of 
eclipse.
} 
\label{fig:eclipseima}
\end{figure}

\begin{figure}
\centering
\includegraphics[angle=-90,scale=0.35]{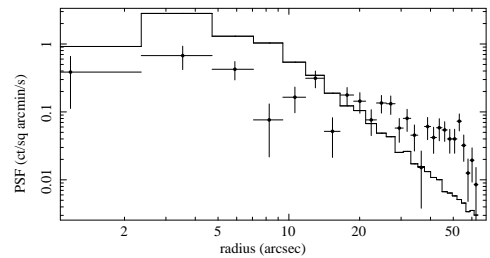}
\caption{Radial distribution of photons  extracted from the image of \src\ during the eclipse 
(0.3-5~keV energy band). The distribution is much flatter than the PSF of a point-like source in 
\swift\,/XRT (solid line).} 
\label{fig:psf}
\end{figure}

To clarify the origin of the residual X-ray emission of \src\ during the eclipse, 
we show in Fig.~\ref{fig:eclipseima} four images of the source extracted 
during the \swift\ observation ID.~00031686002. The two images at the top were accumulated during 
the time interval in which the X-ray source was eclipsed (total exposure time 840~s) and are in the 
0.3--5.0~keV and 5--10~keV energy bands. The source is clearly detected in the lower energy image.  
The other two images were extracted in the same observation and energy bands, but during the 840~s preceding 
the eclipse. These figures suggest that the source was not point-like, but slightly 
extended (by a few arcminutes), especially at lower energies. To support this finding, we compared the 
radial distribution of photons during the eclipse with the point spread function (PSF) from the 
calibrations of \swift\,/XRT (using the tool {\sc ximage}). We report the results in Fig.~\ref{fig:psf}. 
The PSF during the eclipse was much flatter than expected for a point source. 

This indicates that 
the residual emission around the source was most likely due to the effect of a scattering halo 
located halfway to \src\ \citep[see e.g.,][]{day91,predehl95}\footnote{We note that a study of the psf outside 
the eclipse in the observation ID.~00031686002 would also reveal a deviation with respect to the theoretically 
expected profile. However, in this case, this is due to the effect of the pile-up (see Sect.~\ref{sec:swift}).}.  
According to this interpretation, soft X-ray photons emitted when the source is outside the eclipse 
are scattered along our line of sight by interstellar dust, reaching us after
the obscuration of the X-ray source, as a results of the longer path that they
follow. The effect of a dust-scattering halo is most prominent at lower energies, 
because of the energy dependence of the scattering cross-section, 
and when the source is obscured, because direct
photons from the source are then virtually absent.  

We also searched for additional confirmation of our interpretation by comparing the 
spectra of the source during and outside the eclipse. 
By using data in observation ID.~00031686002, we found that the spectrum during the eclipse 
could be described well by an absorbed power-law model with $N_{\rm H}=(3.2^{+0.9}_{-0.8})\times10^{22}$~cm$^{-2}$, 
and $\Gamma$=3.6$\pm$0.6 (C-statistic/d.o.f.=38.4/44, exposure time 840~s). The average 0.5-10~keV flux in the eclipse 
was (1.6$^{+0.1}_{-1.1}$)$\times$10$^{-11}$~erg/cm$^2$/s, with a contribution from  
the photons in the 5-10~keV energy band lower than 20\%. 
This is compatible (to within the errors) with the flux estimated 
during the eclipse by using \rxte\ (see Sect.~\ref{sec:rxte}).  

\begin{figure}
\centering
\includegraphics[angle=-90,scale=0.3]{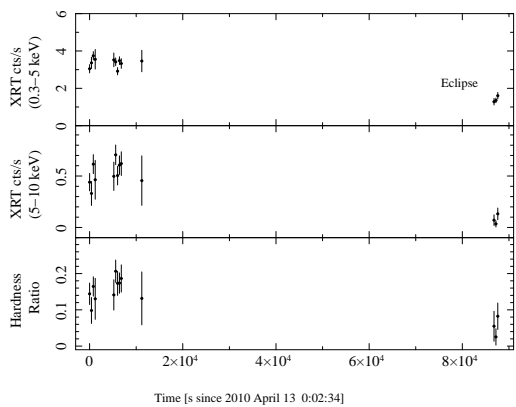}
\caption{\swift\,/XRT light curves of observation 
ID.~00031686002 in the 0.3-5~keV and 5-10~keV energy bands . 
The lower panel shows the corresponding hardness ratio. 
The rates measured during the eclipse are indicated.} 
\label{fig:hratio}
\end{figure}

For comparison, the spectrum of the source extracted outside the eclipse  
is described well by an absorbed power-law model with 
$N_{\rm H}=(3.0\pm0.4)\times10^{22}$~cm$^{-2}$, and $\Gamma$=2.3$\pm$0.2 
($\chi^2_{\rm red}$/d.o.f.=0.99/85, exposure time 2.6~ks). The X-ray flux in this case was 
(2.6$\pm$0.3)$\times$10$^{-10}$~erg/cm$^2$/s, with the photons from the hard energy band (5-10~keV) 
contributing for more than 45\%. 
In Fig.~\ref{fig:hratio}, we also report the light curves of \src\ 
in the observation ID.~00031686002 in two energy bands. 
The hardness ratio, defined as the ratio of the \swift\,/XRT count rate in 
the hard (5-10~keV) to soft (0.3-5~keV) bands versus time, is also shown. Its average value was  
0.16$\pm$0.02 outside the eclipse and 0.04$\pm$0.03 inside the eclipse.  
We conclude that the spectrum during the eclipse was much softer that that outside the eclipse, 
and compatible with the softening by a factor of E$^{-2}$ expected in case the extended emission is 
due to a scattering halo \citep[see e.g.,][and references therein]{day91}.  

We note that, for a halo size $\Theta$ of a few arcmin radius (see Fig.~\ref{fig:eclipseima}) and an 
estimated source distance of $\sim$7~kpc (see Sect.~\ref{sec:source}), the
longer path followed by scattered photons translates into a delay \citep{thompson08} 
\begin{equation}
\delta t = 7.6 d_{\rm 7~kpc} (\Theta/arcsec)^2 x/(1-x) \sim 3\times10^4~s\,,
\end{equation}
where $x$ is the fractional distance of the halo from the source along the line of sight, 
$d_{\rm 7~kpc}$~kpc is the source distance in units of 7~kpc, and we have used $\Theta$=1~arcmin 
and $x$=1/2, as indicative values. The typical delay is thus of few hours, i.e., much longer than 
the duration of the X-ray eclipse ($\sim$2600~s).

\begin{figure*}
\centering
\includegraphics[height=5.5cm]{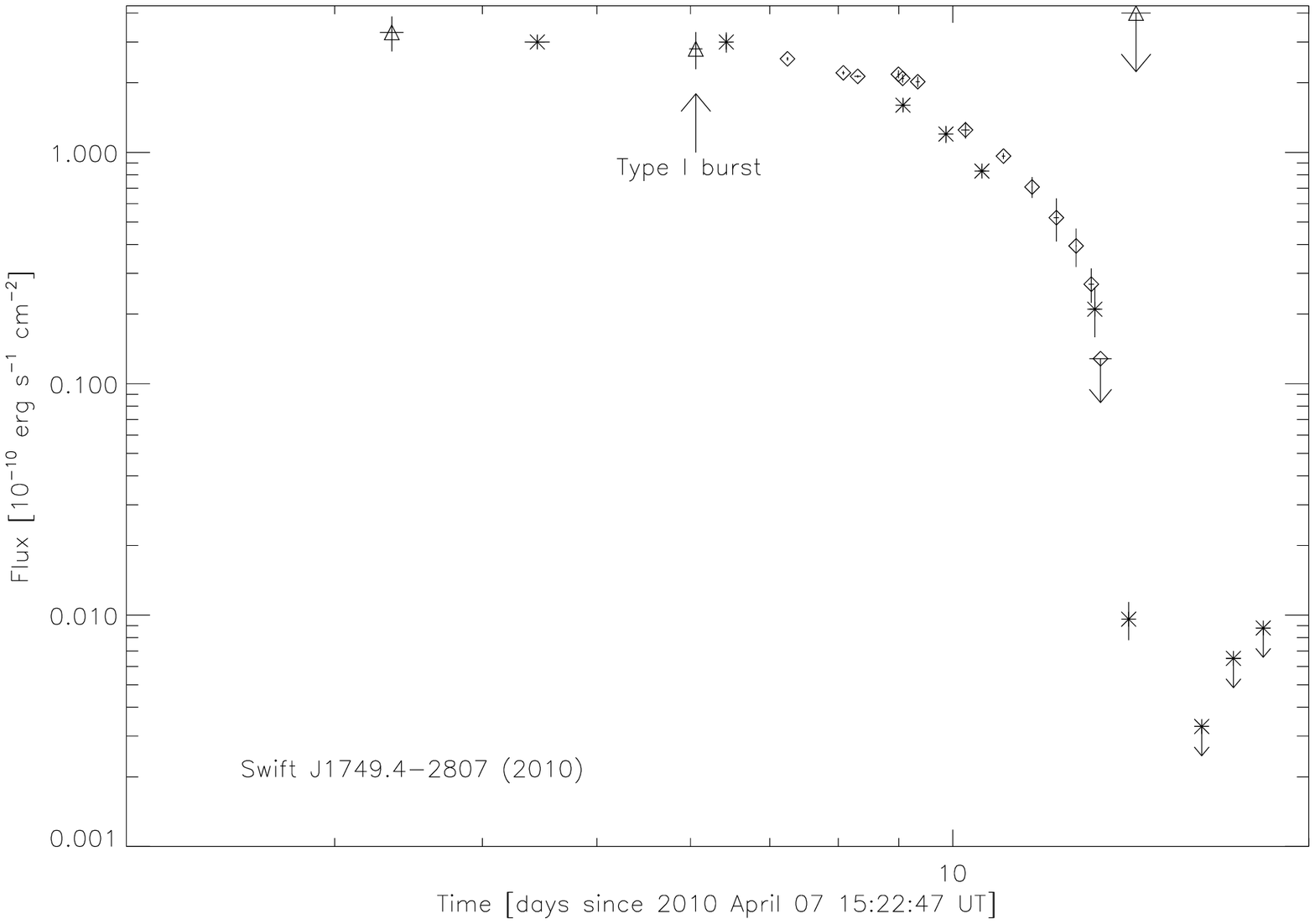}
\includegraphics[height=5.5cm]{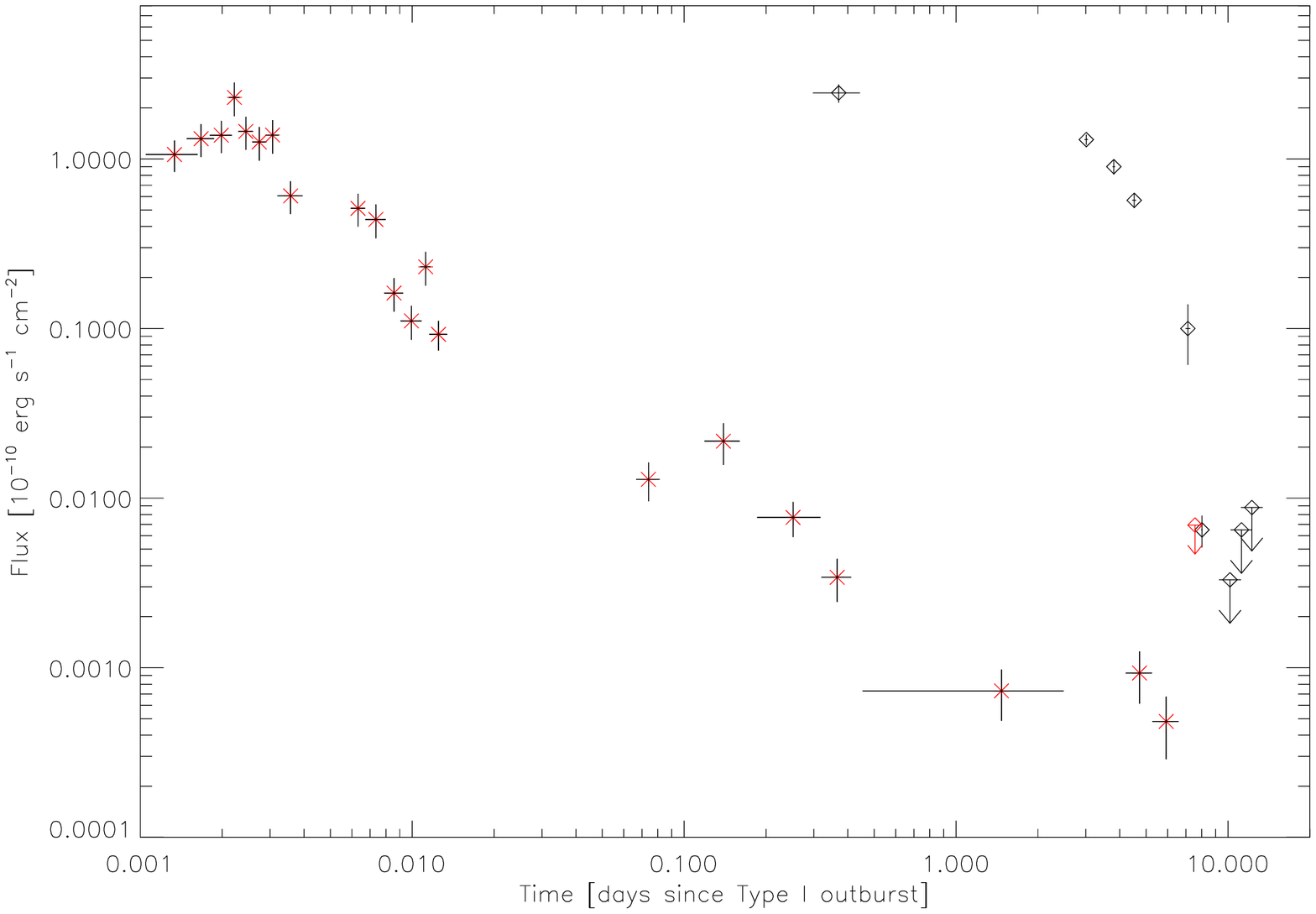}
\caption{{\it Left}: Long-term light curve of the outburst of \src\ in 2010. Stars, 
triangles and diamonds represent respectively \swift,\ \inte,\ and \rxte\  
observations. The downward arrows indicate the upper limits to the source X-ray flux. 
All the fluxes are in the 3-20~keV energy band and have not been 
corrected for absorption. We converted \swift\,/XRT fluxes from the 0.5-10~keV 
energy band to the energy band 3-20~keV by using the spectral model of each observation and the online tool 
{\sc WEBPIMMS} (http://heasarc.nasa.gov/Tools/w3pimms.html). The errors on the fluxes are given at 
90\% c.l. The large upward arrow indicates the time of the type I X-ray burst. 
{\it Right}: The outburst of \src\ occurred in 2006 as observed by \swift\,/XRT (red stars). Here the fluxes are in the 
0.3-10~keV energy band. For comparison, we report in this panel the \swift\,/XRT observation in the same energy 
range carried out during the outburst in 2010 (in black diamonds). The times of the observations in 2006 and 2010 have been scaled to the rise time of the corresponding type-I X-ray burst detected during each event.} 
\label{fig:lcurve}
\end{figure*}

In Fig.~\ref{fig:lcurve}, we report the evolution in the X-ray flux of \src\ during the outburst 
in 2010, derived from the monitoring performed with \inte,\ \rxte,\ and \swift.\ 
The source flux decreased by three orders of magnitude in less than 15 days, displaying a quite clear 
exponential decay. 
A fit with an exponential function to the \swift\ data 
gave an e-folding time of $\tau$=12$^{+7}_{-3}$~days ($\chi^2_{\rm red}$/d.o.f.=15.2/4). 
Even though the $\chi^2_{\rm red}$ is not formally acceptable, we checked that this is due to the relatively 
large scatter in the Swift points during the steep decay. 

In the right panel of Fig.~\ref{fig:lcurve}, we also report for comparison the outburst occurred in 2006, 
observed by \swift\,/XRT. To compare the flux evolution between the two outbursts more reliably, we plot in this panel 
also the \swift\,/XRT observations of the outburst in 2010. In both cases, the times of the observations were scaled  
to the time of the type-I X-ray burst that was detected during each outburst. From this figure it is 
apparent that the outburst in 2010 lasted much longer than the one in 2006 (a factor of $\sim$2), and was characterized 
by a higher averaged X-ray luminosity (the decrease in the X-ray flux with time was, on average, much slower; see also 
Sect.~\ref{sec:discussion}). In both cases, the decrease of the source X-ray flux throughout the outburst 
was not smooth. A relatively large scatter between the fluxes measured from different \swift\ observations, similar to that 
reported above for the outburst in 2010, 
is clearly visible also during the outburst in 2006. \citet{wijnands09} suggested that this scatter 
might be caused by some flares occurring during the decay from the outburst.  

\section{Discussion and conclusions}
\label{sec:discussion}

We have reported on the monitoring of the burst-only 
source \src,\ which was discovered to undergo a new X-ray outburst by \inte\ on 2010 April 10.
 
 \subsection{The outburst decay}
The peak outburst luminosity (see Fig.~\ref{fig:lcurve}) was $\sim$1.8$\times$10$^{36}$~erg/s 
(assuming a distance of 7~kpc), only slightly higher than 
that reported for the outburst that occurred in 2006 ($\sim$10$^{36}$~erg/s). Despite this similarity 
in peak X-ray luminosity, the evolution in the source X-ray flux during the two events was significantly 
different. The 2006 outburst displayed a clear power-law-like decay \citep[index $\sim$-1,][]{wijnands09}, and 
lasted less than $\sim$6~days (see also Sect.~\ref{sec:source}). In contrast, the 2010 outburst lasted 
more that 11~days\footnote{Note that we were unable to 
determine the exact starting point of the outburst, since the first available observation prior to the discovery 
of the renewed activity is on 2010 April 7 and the relatively short exposure time enables to derive only a poorly 
constraining upper limit (see Table~\ref{tab:log}).} and was characterized by an exponential decay 
(see Sect.~\ref{sec:results}). 

Disk instability models for the outbursts of transient NS LMXBs predict that 
both linearly and exponentially decaying outburst can be produced \citep{king98}.   
\citet{king98b} showed that, when the irradiation of the accretion disk from 
the X-ray emission of the NS is strong enough (i.e., the peak X-ray luminosity 
at the onset of the outburst is high), the disk is completely ionized out to its outer  
edge and the light curve of the outburst follows an exponential decay. In this phase, most of the disc 
mass can be accreted onto the NS in a viscous timescale. 
A change in the profile of the outburst is expected once 
the source has faded below a certain luminosity level. Below this threshold, only  
part of the accretion disk can be ionized. As a consequence, both the mass accretion rate onto the NS and the outer boundary 
of the ionized region in the disk begin to decrease with time, leading to a linear (rather than exponential) 
decay \citep{king98,shahbaz98,powell07}. 

The value of the critical X-ray luminosity at which these changes occur 
depends on the properties of the system (total mass and orbital period). 
In the case of \src\,, this luminosity 
is expected to be on the order of 10$^{36}$~erg/s  \citep{shahbaz98}, thus comparable to  
the peak X-ray luminosity estimated during both the events in 2006 and 2010. 
As the event in 2010 was characterized by an X-ray luminosity slightly higher than 
that measured at the peak of the outburst in 2006, it is possible that this relatively 
small difference was sufficient to cause a different behavior during 
the outburst decay. We note that, even though the peak luminosity of the two outbursts 
was almost comparable, during the outburst that occurred in 2010 the source remained 
in a high X-ray luminosity state for a significantly longer time interval than in 2006.  
This might have caused the ionization of a larger portion of the accretion disk and thus led 
to an exponential outburst decay. 
Given the relatively poor number of \swift\ and \rxte\ observations during the 
end of the outburst in 2010, it is not possible to investigate more deeply whether if a switch 
between exponential and linear decay occurred at some point during the steep decay of the 
2010 outburst. 

The discovery of millisecond pulsations in the X-ray emission 
of \src\ secured the previously suggested association of this source (and other burst-only 
sources in general) with the class of the accreting AMSPs. 
This permits a comparison between the two classes.   
In particular, AMSPs are known to undergo week-to-month-long outbursts that usually 
exhibit an initial exponential decay, followed by a linear one,
in agreement with the prediction of the disk instability model discussed above. 
An in-depth study of the outburst decay of a number of AMSPs was carried out by \citet{powell07}. These authors 
showed that most of these sources display a ``knee'' in their decay, 
when the X-ray luminosity  
falls below the threshold at which the switch between exponential to linear decay is predicted. 

This behavior was observed in the AMSPs SAX\,J1808.4$-$3658, XTE\,J1751$-$305, and XTE\,J0929$-$314. 
Some anomalies in the outburst decay were reported for the AMSPs SAX\,J1808.4$-$3658, XTE\,J1807$-$294, 
and IGR\,J1751$-$305. During an outburst in 2000, SAX\,J1808.4$-$3658 displayed several unexpected 
``re-flares'' about $\sim$20~days after the initial exponential+linear decay \citep{wijnands06,campana08}, 
whereas the source XTE\,J1807$-$294 underwent in 2003 a rather long ($\gtrsim$100~day) pure exponential 
outburst without any clear evidence of a linear phase \citep[see e.g.,][and references therein]{falanga2005,powell07}. 
Another AMSP, IGR\,J1751-305, displayed a few outbursts lasting exceptionally only a few days and 
reaching a relatively low X-ray luminosity \citep[a few mCrab, see e.g.,][]{falanga07,markwardt07}. 
The outbursts observed from \src\ in 2006 and 2010 were both significantly shorter than those observed from 
the other AMSPs, and only the anomalous outbursts detected from IGR\,J1751-305 resemble those 
of \src\ .

Similar dim outbursts have been observed from the very faint X-ray transients 
(VFXTs, see Sect.~\ref{sec:intro}), and an association between these sources 
and the burst-only sources was suggested by \citet{campana09}. Different ways of interpreting 
the behavior of the VFXTs were reported by \citet{king06}. These authors showed that the 
VFXTs hosting NS primaries can be relatively well understood if the material accreted on these systems 
originates from brown dwarfs or planetary companions. However, in the case of \src\ such 
low-mass companions is ruled out by the mass function of the system, which implies a minimum companion 
mass of $\sim$0.5$M_{\odot}$ assuming a NS of  $1\,M_{\odot}$.  

\subsection{Timing features}
In the case of \src\,, an in-depth study of the properties of the primary and secondary stars  
is possible for the first time by using the orbital solution and characteristics of the X-ray 
eclipses. At present, \src\ is the only eclipsing AMSPs. An extensive discussion of these issues
is beyond the scope of the present 
work, and will be reported in a separate paper (Campana et al., 2010, in preparation).  

In Sect.~\ref{sec:halo}, we carried out a careful analysis of the residual X-ray 
emission of \src\ during the eclipse. While our analysis of the \rxte\ data could not unambiguously  
identify the origin of this emission, we found in the \swift\ data strong evidence that residual 
emission during the eclipse is caused by a scattering halo 
in the direction of the source. The effect of a dust scattering halo is 
most prominent when a source is in eclipse and at lower energies ($\lesssim$3~keV, 
see Sect.~\ref{sec:halo}). Since \src\ is the first eclipsing AMSP, it is also the first source 
in this class for which this effect could be detected. 

An additional peculiarity of \src\ was revealed by the Fourier analysis of the \rxte\ data, 
which showed a signal at the spin frequency of the NS and its second harmonic.   
This is consistent with the fact that the pulse profiles clearly displayed 
a double peak (see Fig.~\ref{fig:timepulses}). This feature has never been observed in another AMSP and 
is most likely due to emission from both accreting polar caps that sweeps across our line of sight.
This geometry is possible because the system is seen  
nearly edge-on (as confirmed by the detection of X-ray eclipses). Moreover, the study of the pulse profile 
in different \rxte\ observations has found that the shape and amplitude of the 
two peaks changed significantly during the outburst decay. We propose that this behavior is caused by 
changes in the geometry of the emitting region close to the NS surface  and/or changes in the
radius and/or thickness of the inner edge of the accretion disk.

\subsection{Type I X-ray burst}
Our analysis of the data presented here has also detected a type-I X-ray burst 
in the JEM-X2 light curve. This is the second type-I X-ray burst reported so far from \src.\  
The first burst was discovered with \swift\,/BAT \citep{wijnands09} and was characterized by a peak flux 
a factor of $\sim$2 higher than that measured by JEM-X2 (see also Sect.~\ref{sec:source}).  
The statistics of the JEM-X2 were of too low quality to perform a time-resolved spectral analysis of the burst,  
and no evidence of a photospheric radius expansion could be found (see Sect.~\ref{sec:results}). 
The observation of a type-I X-ray burst can be used to derive an upper 
limit to the source distance by assuming that the peak X-ray luminosity of the burst corresponded to the 
Eddington value $L_{\rm Edd}$$\approx$3.8$\times$10$^{38}$~erg/s  
\citep[as empirically derived by][for a helium burst]{kul03}. However, since the peak flux of the JEM-X2 
burst was significantly lower than that observed previously by \swift,\ the latter already provided the most 
restrictive upper limit\footnote{We note however that, if a pure hydrogen burst is considered, then the inferred upper 
limit to the source distance would be $\approx$5~kpc.} of 7~kpc on the distance to \src\ (see Sect.~\ref{sec:intro}). 
Using this distance, the persistent unabsorbed 0.1-100~keV flux of the source at the time of the JEM-X2 burst 
would translate into a bolometric luminosity of $L_{\rm pers}\approx~4.7\times 10^{36}$~erg/s, or 1.2\% $L_{\rm Edd}$. 
This corresponds to a local accretion rate per unit area of $\dot{m}\approx$3.2$\times$10$^{3}$~g/s/cm$^2\simeq$1.2\%$\dot m_{\rm Edd}$,  
where we have used the relation $L_{\rm pers}=4\pi R^2\dot m(GM/R)/(1+z)$ 
\citep[with $z$=0.31 the NS gravitational redshift; see e.g.,][]{lewin93}. 

Theoretical models predict that, when 
0.01$<$$\dot{m}$/$\dot{m_{\rm edd}}$$<$0.1, H burns stably through the hot CNO cycle 
and a layer composed of pure He develops underneath the NS surface. This layer can then 
ignite by means of the 3$\alpha$ process and lead to a pure He burst, with a typical rise time 
of $\sim$1~s and a duration of $\sim$10~s.   
These quantities, estimated for the burst 
detected by JEM-X2 (see Sect.~\ref{sec:results}),
are fully compatible with those expected for a pure He burst.

We note that pure He runaways were also reported for the other burst-only source 
GRS\,1741.9-2853 \citep{trap09}. The detection of this kind of type-I X-ray bursts 
from the burst-only sources is particularly intriguing because it would  
argue against the idea that these sources are the prototypes of the (poorly 
observed) H-burning bursts with low accretion rates 
\citep[see Sect.~\ref{sec:intro} and, e.g.][]{peng07}. 

Finally, we can check the consistency of the results derived above for the type-I X-ray burst 
by evaluating the theoretically expected recurrence time of the burst. 
We first estimate the ignition depth of the burst, $y_{\rm ign}$, 
using the equation $E_{\rm burst}=4\pi R^2y_{\rm ign}Q_{\rm nuc}/(1+z)$, where   
$E_{\rm burst}=4\pi d^2f_b=2.0\times10^{39}$ erg ($d$/7 kpc), $f_b$ is the measured fluence of the burst 
(see Table~\ref{tab:burst}), and  $Q_{\rm nuc}\approx 1.6$ MeV corresponds to the nuclear
energy release per nucleon for complete burning of helium to iron group elements 
\citep{wallace81,fujimoto87}. We obtained $y_{\rm ign}$=$1.4\times$10$^8$~${\rm g\ cm^{-2}}$.  
For the above values of the local accretion rates and ignition depth, the expected recurrence time of a 
He bursts is about $\Delta$t=($y_{\rm ign}$/$\dot m$)$(1+z)$$\simeq$0.6~days 
(independent of the assumed distance). The burst detected by JEM-X2 
occurred $\sim6.8$~d after the first available \inte\ observation (see Fig.~\ref{fig:lcurve}). 
However, the total effective exposure time on the source was $\sim$0.9~days, thus compatible 
with the observation of a single type-I X-ray burst throughout the observational period.\\

At the time of writing this work, 
two other papers were submitted describing the same source: \citet{markwardt10b} and \citet{altamirano2010b}.  
These authors focused mostly on the timing analysis of the system
and discussed in detail the dynamical constraints on the binary
system that could be derived by using the eclipses detected in the RXTE data. 
In our paper, we have concentrated more on the study and interpretation of the evolution of the timing and spectral properties of the source during the outburst. 
We checked (where possible) that the results derived in our paper are fully compatible with those reported in the other two papers. We have also discussed for the first time the properties of the eclipse found in the \swift\ data, and provided an interpretation of the residual flux during the eclipse.

\section*{Acknowledgments}
EB and CF thank N. Gehrels and the \swift\ team for their availability
and prompt response in carrying out follow-up observations of \src. We thank
S. Suchy for support from the HEXTE instrument team and the research groups of IAA-T\"ubingen and
Dr. Remeis-Sternwarte in Bamberg for developing and making available useful scripts to analyze \rxte\ data. 
This research has made use of the XRT Data Analysis Software (XRTDAS) 
developed under the responsibility of the ASI Science Data Center 
(ASDC), Italy. AP acknowledges financial support from the Autonomous Region of
Sardinia through a research grant under the program PO
Sardegna FSE 2007Ð2013, L.R. 7/2007 ``Promoting scientific
research and innovation technology in Sardinia''.

\end{document}